\documentclass[twocolumn,a4]{article}

\usepackage{graphicx}             
\usepackage{subfigure}            
\usepackage{url}                  
\usepackage[T1]{fontenc}
\usepackage{nicefrac}             
\usepackage{indentfirst}          
\usepackage[super, negative]{nth}
\usepackage{amsfonts}
\usepackage{amssymb}
\usepackage{multirow}
\usepackage{balance}

\usepackage[dvipsnames,usenames]{color}
\usepackage{xspace}
\usepackage{array,booktabs}
\usepackage{latexsym}
\usepackage{pifont}
\usepackage{colortbl}
\usepackage{bytefield}

\usepackage{algorithm}
\usepackage[noend]{algpseudocode}

\usepackage{xcolor}

\usepackage{authblk}

\definecolor{my-gray}{gray}{0.45}
\algnewcommand{\ccomment}[1]{\State \textit{/* #1 */}}
\algnewcommand{\icomment}[1]{\Comment{\textit{\color{gray}#1}}}
\algnewcommand{\iccomment}[1]{\textit{\color{my-gray}{~/* #1} */}}

\algnewcommand\Or{~\textbf{or}~}
\algnewcommand\And{~\textbf{and}~}
\algnewcommand\In{~\textbf{in}~}

\newcommand{\dfn}[1]{\textit{#1}}            

\newcommand{\ttlexceeded}{\texttt{time-exceeded}\xspace}

\newcommand{\dstunreach}{\texttt{destination-unreachable}\xspace}
\newcommand{\msglong}{\texttt{message-too-long}\xspace}
\newcommand{\traceroute}{\texttt{traceroute}\xspace}

\newcommand{\copycat}[0]{\texttt{copycat}\xspace}

\graphicspath{{.}{./Pictures/}}

\begin{document}






%

\title{Using UDP for Internet Transport Evolution}



\author[1]{Korian Edeline}
\author[2]{Mirja K\"uhlewind}
\author[2]{Brian Trammell}
\author[3]{Emile Aben}
\author[2]{Benoit Donnet}
\affil[1]{Montefiore Institure, Universit\'e de Li\'ege, Belgium}
\affil[2]{Networked Systems Group, ETH Z\"urich, Switzerland}
\affil[3]{RIPE NCC, Amsterdam, Netherlands}

\date{ETH TIK Technical Report 366, 22 December 2016}

\maketitle

\begin{abstract}

The increasing use of middleboxes (e.g., NATs, firewalls) in the Internet has
made it harder and harder to deploy new transport or higher layer protocols,
or even extensions to existing ones. Current work to address this Internet
transport ossification has led to renewed interest in UDP as an encapsulation
for making novel transport protocols deployable in the Internet. Examples
include Google's QUIC and the WebRTC data channel. The common assumption made
by these approaches is that encapsulation over UDP works in the present
Internet. This paper presents a measurement study to examine this assumption,
and provides guidance for protocol design based on our measurements.

The key question is ``can we run new transport protocols for the Internet over
UDP?'' We find that the answer is largely ``yes'': UDP works on most networks,
and impairments are generally confined to access networks. This allows
relatively simple fallback strategies to work around it. Our answer is based
on a twofold methodology. First, we use the RIPE Atlas platform to basically
check UDP connectivity and first-packet latency. Second, we deploy \copycat, a
new tool for comparing TCP loss, latency, and throughput with UDP by
generating TCP-shaped traffic with UDP headers.

\end{abstract}


\section{Introduction}\label{intro}
Most Internet applications today are built on top of the Transport Control
Protocol (TCP), or some session-layer protocol that uses TCP, such as the
Hypertext Transfer Protocol (HTTP) or WebSockets. Indeed, the ubiquity and
stability of TCP as a common facility that handles the hard problems of
reliability and congestion control is a key factor that has led to the massive
growth of the Internet.

However, not every application benefits from the single-stream, fully-reliable
service provided by TCP. In addition, the ubiquitous deployment of network
address translators (NATs) and firewalls that only understand a limited set of
protocols make new protocols difficult to deploy. Previous attempts to deploy
new protocols such as the Stream Control Transmission Protocol
(SCTP)~\cite{rfc4960} were hindered by this ossification~\cite{natSCTP}, as well
as by the difficulty of rapid deployment of new kernel code across multiple
platforms. The deployment of middleboxes that ``understand'' TCP also limit the
ability to deploy new TCP options and features~\cite{TCPExposure}. Much of the
design work in Multipath TCP~\cite{rfc6182,rfc6824}, for example, addressed
middlebox detection and avoidance.

This has led to a current trend in transport protocol design to use UDP
encapsulation to solve this problem. Google's \dfn{Quick UDP Internet
Connections} (QUIC)~\cite{draft-hamilton-early-deployment-quic-00} and the
WebRTC data channel~\cite{draft-ietf-rtcweb-data-channel} both use UDP as an
``outer'' transport protocol. In both cases the transport protocol dynamics
(connection establishment, reliability, congestion control, and transmission
scheduling) are handled by the ``inner'' protocol. In the case of the WebRTC
data channel, this is \dfn{SCTP over Datagram Transport Layer Security}
(DTLS)~\cite{rfc6083}; in the case of QUIC, it is the QUIC transport protocol
itself. This is a new kind of encapsulation. In contrast to traditional
tunneling, this approach borrows the ``wire image'' of UDP for two key benefits:
userspace deployability of new transports, due to the ubiquitous availability of
UDP sockets for unprivileged, userspace programs; and NAT/firewall traversal, as
most such devices recognize UDP ports. The \dfn{Path Layer UDP Substrate} (PLUS)
effort within the IETF~\cite{draft-trammell-plus-abstract-mech,draft-trammell-plus-statefulness}\footnote{Note that PLUS is
an evolution and change in scope from the previous Substrate Protocol for User
Datagrams (SPUD) effort within the IETF; PLUS shares most of its requirements~\cite{draft-trammell-spud-req}
with SPUD.} generalizes this approach for new transport protocols.

This work presents a measurement study aimed at evaluating the correctness of
the assumption underlying these approaches: that such UDP encapsulation will
work in the present Internet, and that connectivity and performance of UDP
traffic are not disadvantaged with respect to TCP based only on the presence
of a UDP header. We do so in two ways. First, we measure UDP connectivity and
first-packet latency with the RIPE Atlas measurement platform from a wide
variety of vantage points, to get information about basic UDP blocking on
access networks. Second, we use a novel measurement tool called \copycat to
create TCP traffic with UDP's wire image, and perform full-mesh measurements
on a wide variety of test networks: PlanetLab, Ark~\cite{ark}, and cloud
service provider Digital Ocean~\cite{digital_ocean}, in order to determine if
differential treatment of UDP and TCP packets might disadvantage 
congestion-controlled traffic with UDP headers.

These measurements are important because we know of several ways in which these
assumptions may not hold. UDP is blocked by firewall rules on some restrictive
access networks, especially within enterprises~\cite{Reddy2015}.  In addition to
complete blocking, other impairments to traffic with UDP headers may exist, such
as throttling or fast NAT timeouts. There are also implicit first-party claims
by at least one major mobile operator that it rate-limits UDP traffic as a
preemptive defense against distributed denial of service 
attacks~\cite{draft-byrne-opsec-udp-advisory}.

In summary, we see evidence of complete blocking of UDP in between about $2\%$
and $4\%$ of terrestrial access networks, and find that blocking is primarily
linked to access network; these results are in line with reported QUIC
performance~\cite{quic-performance-google}. We note these networks are not
uniformly distributed throughout the Internet: UDP impairment is especially
concentrated in enterprise networks and networks in geographic regions with
otherwise- challenged connectivity. Where UDP does work on these terrestrial
access networks, we see no evidence of systematic impairment of traffic with
UDP headers. The strategy taken by current efforts to encapsulate new
transports over UDP is therefore fundamentally sound, though we do give some
guidance for protocol design that can be taken from our measurements in
Sec.~\ref{conclusion}.

\section{Related Work}\label{related}

We present the first attempt to directly evaluate performance differences of
congestion-controlled, reliable traffic based solely on the wire image (i.e.,
whether a TCP or UDP header is seen on the traffic by the network). This
measurement study complements a variety of measurement works past and present.

Google's deployment of its QUIC protocol, of course, represents a much larger
scale experiment than that presented here, but it is limited to a single
content provider's network. Google has reported results from this experiment,
but only in highly aggregated form~\cite{ietf96_quic}: of users of Chromium
selected for QUIC experimentation connecting to Google web properties on UDP
port 443, 93\% of connections use QUIC. 2\% use TCP because TCP has lower
initial latency. In 5\% of cases, UDP is impaired: either rate-limited
(0.3\%), blocked at the access network border (0.2\%), or blocked on the
access network close to the user (4.5\%). Google reports a downward trend in
UDP rate limiting during the course of the experiment. QUIC has been measured
in controlled environments outside Google, as well: Carlucci et
al~\cite{Carlucci2015} compare HTTP/2 performance over QUIC with TCP, presuming
an unimpaired network.

Related to differential treatment is differential NAT and firewall state
timeout. H\"at\"onen et al~\cite{Haetoenen2010} looked at NAT timeouts on a
variety of 34 home gateway devices available in 2010, and found a median idle
timeout for bidirectional UDP traffic of about 3 minutes, with a minimum
timeout of 30 seconds. In contrast, median idle timeout for TCP was 60
minutes, with a minimum of 5 minutes.

Network- and transport-layer performance in general is a well-studied field:
Qian et al.~look at the characteristics of measured TCP
traffic~\cite{tcp_revisited}. Paxson et al.~\cite{Paxson1997} focuses on packet
dynamics of TCP bulk transfers between a limited set of Internet sides.
Pahdye et al.~\cite{Pahdye2001} investigates TCP behavior of web server,
assuming no interference in the network. Xu at al.~\cite{Xu2014} uses UDP-based
traffic to evaluate characteristics of cellular networks. They also test TCP
throughput to ensure that no UDP throttling was performed in the tested network
that would tamper their results. Melia et al.~evaluate TCP and UDP performance
in an IPv6-based mobile environment~\cite{handover}.
Sarma evaluates TCP and UDP performance through simulation in a particular
context (QoS) considering two queuing mechanisms: RED and Drop
Tail~\cite{tcp_udp_queue}. Bruno et al develop models for analyzing and
measuring UDP/TCP throughput in WiFI networks~\cite{throughput_802.11}. While
some of these results provide insights and background knowledge on aspects of
UDP as well as TCP performance, they can not be used to answer the question of
differential treatment in the Internet (covering different access network
technologies) that we ask in this paper.

Network measurement tools have been proposed to evaluate reachability (e.g.,
Netalyzr~\cite{netalyzr} determines whether a particular service, identified
by its port number and transport protocol, is reachable) or transport protocol
performance and analysis (e.g., iPerf~\cite{iperf}, tbit~\cite{tbit}).  
These tools, however, are not designed to 
measure differential treatment between UDP and TCP. Packet encapsulation 
for network measurements as employed by \copycat is a common technique, 
as well, particularly for middlebox identification.
For instance, the TCPExposure~\cite{TCPExposure} client sends TCP packets over
raw IP sockets toward a user-controlled server.  The server sends back
received and to-be-sent headers as payload so that the client can compare what
was sent to what was received.

\section{Measurement Methodology}\label{methodology}

This study uses two separate methodologies to reach its conclusions:
differential treatment measurement using \copycat
(Sec.~\ref{methodology.copycat}) and connectivity and latency comparison using
RIPE Atlas (Sec.~\ref{methodology.atlas}).

\subsection{Copycat}\label{methodology.copycat}
\begin{figure}[!t]
  \begin{center}
    \includegraphics[width=8cm]{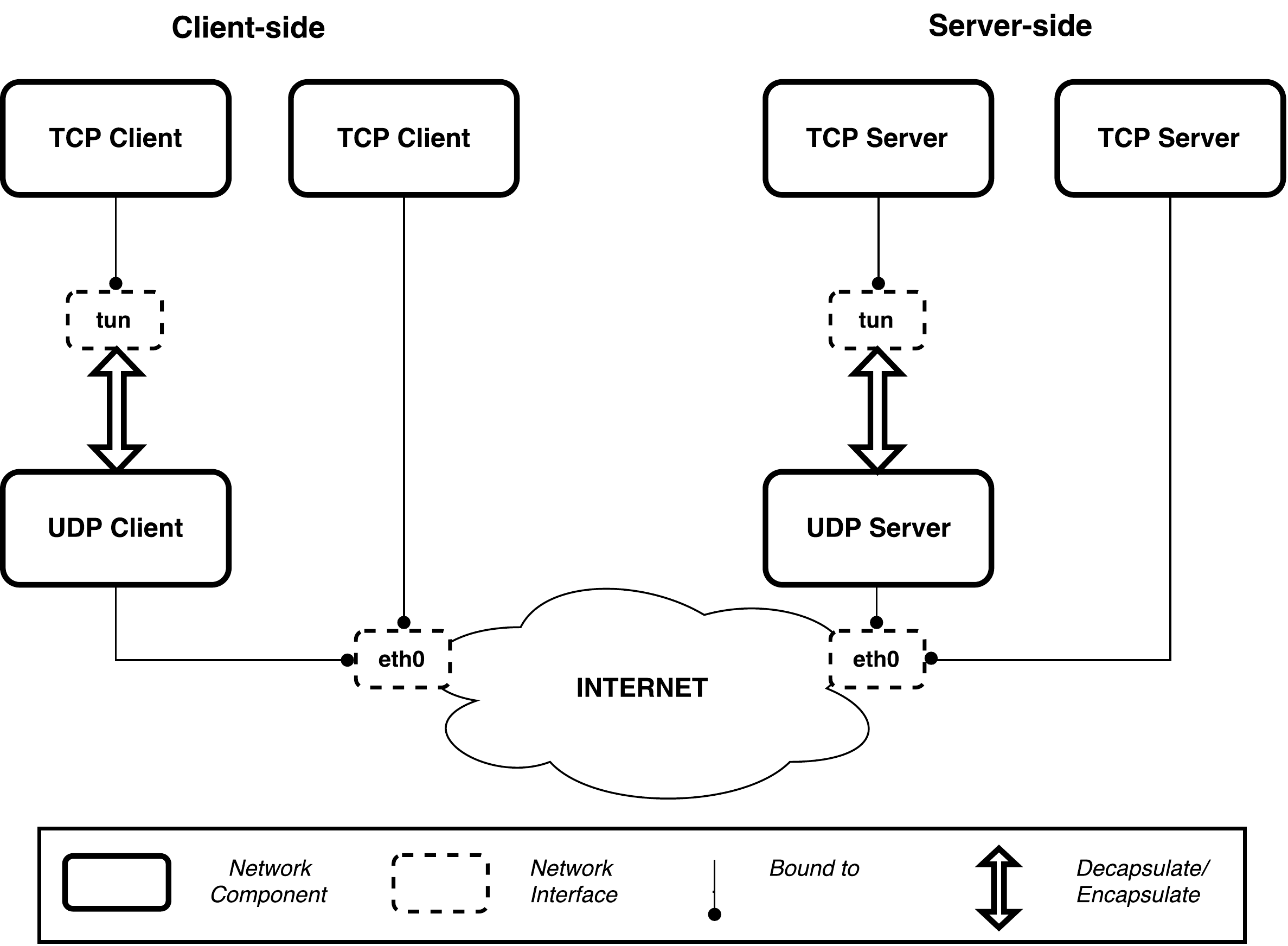}
  \end{center}
  \caption{\copycat measurement methodology.}
  \label{methodology.copycat.fig}
\end{figure}

\copycat\footnote{Sources are freely available at 
\url{https://github.com/mami-project/udptun}}
simultaneously runs pairs of flows between two
endpoints. One is a normal TCP flow and one a TCP flow using UDP as an
``outer'' transport. This allows us to evaluate differences in connectivity
and quality of service due to differential network treatment based solely on
transport protocol header. This TCP flow is used to simulate a new transport
running over UDP, by providing traffic with TCP- friendly congestion control.
The two flows run in parallel with the exact same 4-tuples, to obtain flows
with the most similar possible treatment from the network, but with different
transport headers. By comparing performance of these flows to each other, we
are able to isolate differences that can be attributed to differential
treatment by the path.

As shown in Fig.~\ref{methodology.copycat.fig}, the UDP flow is obtained by
tunneling a TCP flow over UDP. To achieve this, \copycat first creates a
\texttt{tun} virtual network interface that simulates a network layer device
and operates at Layer 3. In our measurement setup, each node runs both the
\copycat client and the server. On the client side, the TCP client connects
to its peer via the Internet-facing interface and receives data from it,
writing it to disk. The UDP client consists of the TCP client bound to the
\texttt{tun} interface, which is in turn bound by \copycat to a UDP socket on
the Internet-facing interface. \copycat thus works as a tunnel endpoint,
encapsulating TCP packets from \texttt{tun} in UDP headers, and decapsulating
received UDP packets back to TCP packets to \texttt{tun}. The server-side
consists of a similar arrangement, listening for connections from clients and
sending data to them. The client waits for both transfers, via TCP and TCP-
controlled UDP, to be completed before connecting to the next destination.


Each flow consists of a unidirectional data transfer. The smallest flow is
calibrated not to exceed the TCP initial window size, which range from $2$-$4$
to $10$ Maximum Segment Size (MSS) depending on the kernel version used by the
different measurement platforms~\cite{RFC3390, RFC6928}. This ensures that in
the smallest flow, we send all data segments at once. Then, we increase the
size of the flows by arbitrary factors of $3$, $30$, $300$, and $1500$ to
observe the impact of differential treatment for congestion-controlled traffic
with larger flows. 



To avoid unwanted fragmentation of UDP datagrams and ICMP \msglong errors, and
to ensure that packets from both tunneled and non-tunneled flows are equally
sized, we decrease the MSS of the tunneled TCP flow by the size of the tunnel
headers (IP header + UDP header = 28 Bytes).

\copycat is coded in C to minimize overhead. I/O multiplexing
is handled using \texttt{select()}. All network traces are captured at
\texttt{eth0} using \texttt{libpcap}.

We deployed \copycat on the PlanetLab distributed testbed on the entire pool
(153) of available nodes between March \nth{6} and April \nth{23}, 2016.
Considering PlanetLab port binding restrictions (e.g., 80, 8000, and 53, 443 on
certain nodes), we chose seven ports-53, 443, 8008, 12345, 33435, 34567, and
54321- respectively DNS, HTTPS, HTTP alternate, a common backdoor, the Atlas UDP
traceroute default, an unused and an unassigned port, to maximize routers policy
diversity. For each port and pair of nodes, we generated 20 pairs of flows of
$1$, $3$, and $30$ TCP initial windows, and 10 pairs of flows of $300$ and
$1,500$ TCP initial windows, for a total of 4,908,650
flows.\footnote{\label{ft_dataset}The complete dataset is freely available at
\url{http://queen.run.montefiore.ulg.ac.be/~edeline/copycat/}.}

Then, we selected 93 nodes (one per subnetwork) from the entire pool to maximize 
path diversity. The selected nodes are located in 26 countries across 
North America (44), Europe (29), Asia (13), Oceania (4), and South America (3).
The filtered PlanetLab dataset then consists in 1,634,518 flows.

We also deployed \copycat on $6$ Digital Ocean nodes,
located in 6 countries across North America (2), Europe (3), and Asia (1). Given 
the less restrictive port binding policies and the more restrictive bandwidth
occupation policies, we tested ports 80 and 8000 in addition of the PlanetLab
ports. For each port, we generated 20 pairs of flows of $1$, $3$, and $30$ 
TCP initial windows size between May \nth{2} and \nth{12}, 2016. We repeated
the same methodology for both IPv4 and IPv6. This dataset consists in 32,400
IPv4 and 31,366 IPv6 flows.

\subsection{RIPE Atlas}\label{methodology.atlas}
\begin{table*}[!htbp]
  \begin{center}
    \begin{tabular}{cl|cc|c}   
     \multicolumn{2}{c|}{\multirow{3}{*}{\textbf{Dataset}}} & \multicolumn{3}{c}{\textbf{Results}}\\
     & & \multicolumn{2}{c|}{\# Probes} & \multirow{2}{*}{No UDP Connectivity} \\
     & & total & failed & \\
       \hline
       \multirow{5}{*}{RIPE Atlas} & Latency, 2015 & 110 & 0 & 0.00\% \\
        \cline{2-5}
        & All UDP, 2015 & 2,240 & 82 & 3.66\% \\
        \cline{2-5}
        & all MTU, March 2016         & 9,262 & 296 & 3.20\% \\
        & \hspace{2mm}72 bytes        & 9,111 & 244 & 2.68\% \\
        & \hspace{2mm}572 bytes       & 9,073 & 210 & 2.31\% \\
        & \hspace{2mm}1,454 bytes     & 8,952 & 137 & 1.53\% \\
        \hline
        \multirow{3}{*}{\copycat}& PlanetLab        & 30,778 & 825 & 2.66\% \\
                                 & Digital Ocean v4 & 135    & 0   & 0.00\% \\
                                 & Digital Ocean v6 & 135    & 0   & 0.00\% \\
    \end{tabular}
  \end{center}
  \caption{Overview of our results on UDP connectivity.  The upper part of the
  table shows the percentage of probes with UDP being blocked, as measured by
  RIPE Atlas in 2015 and 2016 (Sec.~\ref{results:blocking} for details).  The
  lower part shows UDP blocking as measured by \copycat.}  
  \label{results.tab}
\end{table*}

We used the RIPE Atlas~\cite{atlasRipe} measurement network to provide another
view on connectivity and first-packet latency differences between UDP and TCP,
as well as to investigate UDP blockage on access networks and possible MTU
effects on such UDP blockage.

First, we compared latency to last hop from Atlas UDP \traceroute and TCP
\traceroute measurements from a set of 115 Atlas probes in 110 networks
(identified by BGP AS number) to 32 Atlas anchors (i.e., Atlas nodes having
higher measurement capacities than standard Atlas probes), and used this as a
proxy metric for first-packet latency for UDP or TCP. To review, TCP \traceroute
sends \texttt{SYN}s with successively increasing TTL values and observes the
ICMP \ttlexceeded responses from routers along the path and the \texttt{SYN+ACK}
or \texttt{RST} from the target. UDP \traceroute sends packets to a UDP port on
which presumably nobody is listening, and waits for ICMP \ttlexceeded or
\dstunreach responses from the path and target respectively. We set the initial
TTL to 199, which is sufficient to reach the destination in one run without
generating any \ttlexceeded messages from the path, i.e., treating \traceroute
as a simple ping. The measurements ran between September \nth{23} and \nth{26},
2015, with all probes testing each anchor sequentially, sending three packets in
a row once every twenty minutes, for up to 17 connection attempts (51 packets)
each for  UDP/33435, TCP/33435, and TCP/80.

Second, while ``common knowledge'' holds that some networks severely limit or
completely block UDP traffic, this is not the case 
on any of the selected probes in the first
measurement. To get a handle on the prevalence of such UDP-blocking networks,
we looked at 1.1 million RIPE Atlas UDP \traceroute measurements run in 2015,
including those from our first campaign. Here, we assume that probes which
perform measurements against targets which are reachable by other probes using
UDP traceroutes, but which never successfully complete a UDP \traceroute
themselves, are on UDP blocked access networks.

Third, we used a single campaign of about 2.5 million UDP and ICMP traceroutes
from about 10,000 probes with different packet sizes in March 2016, to compare
protocol-dependent path MTU to a specific RIPE Atlas anchor. We compared
success rates with UDP at different packet sizes to ICMP.

\section{Results}\label{results}


Tables~\ref{results.tab} and~\ref{results.table.bias} provide an overview on our main results. In summary,
we show that, aside from blocking of UDP on certain ports, as well as
relatively rare blocking of all UDP traffic on about one in thirty access
networks, UDP is relatively unimpaired in the Internet. 
We explore the details of both tables and additional measurement results in the subsections below.


\subsection{Incidence of UDP Blocking}\label{results:blocking}
Of the 2,240 RIPE Atlas probes which performed UDP \traceroute measurements
against targets which were reachable via UDP \traceroute in 2015, 82 (3.66\%)
never successfully completed a UDP \traceroute. We take this to be an
indication that these probes are on UDP-blocking networks. The location of the
blockage, determined by the maximum path length seen in a UDP \traceroute, is
variable, with the median probe seeing at least one response from the first
five hops. These UDP-blocked probes are more likely than the population of all
examined probes to be on networks in sub-saharan African and east Asian
countries.

Our investigation of MTU issues showed no significant relationship between
packet size and probe reachability up to 1,420 bytes per packet, as compared
to ICMP. In this shorter study in March 2016 using more probes, 296 of 9,262
probes (3.20\%) did not receive a response from the target from the UDP
\traceroute for any packet size. For 72, 572, and 1,454 byte packets,
respectively, 2.68\%, 2.31\%, and 1.53\% of probes received no response to a
UDP \traceroute attempt when receiving a response from an ICMP \traceroute of
the same size. These results are summarized in Table~\ref{results.tab} (upper
part of the table). Note that the relative UDP blocking numbers go down as the
packet size goes up; this is because large ICMP packets are more often blocked
than large UDP packets. From these results, we conclude that differential
treatment between UDP and TCP should not pose a challenge to using UDP as an
outer transport.




\begin{figure}[!t]
  \begin{center}
    \includegraphics[width=8cm]{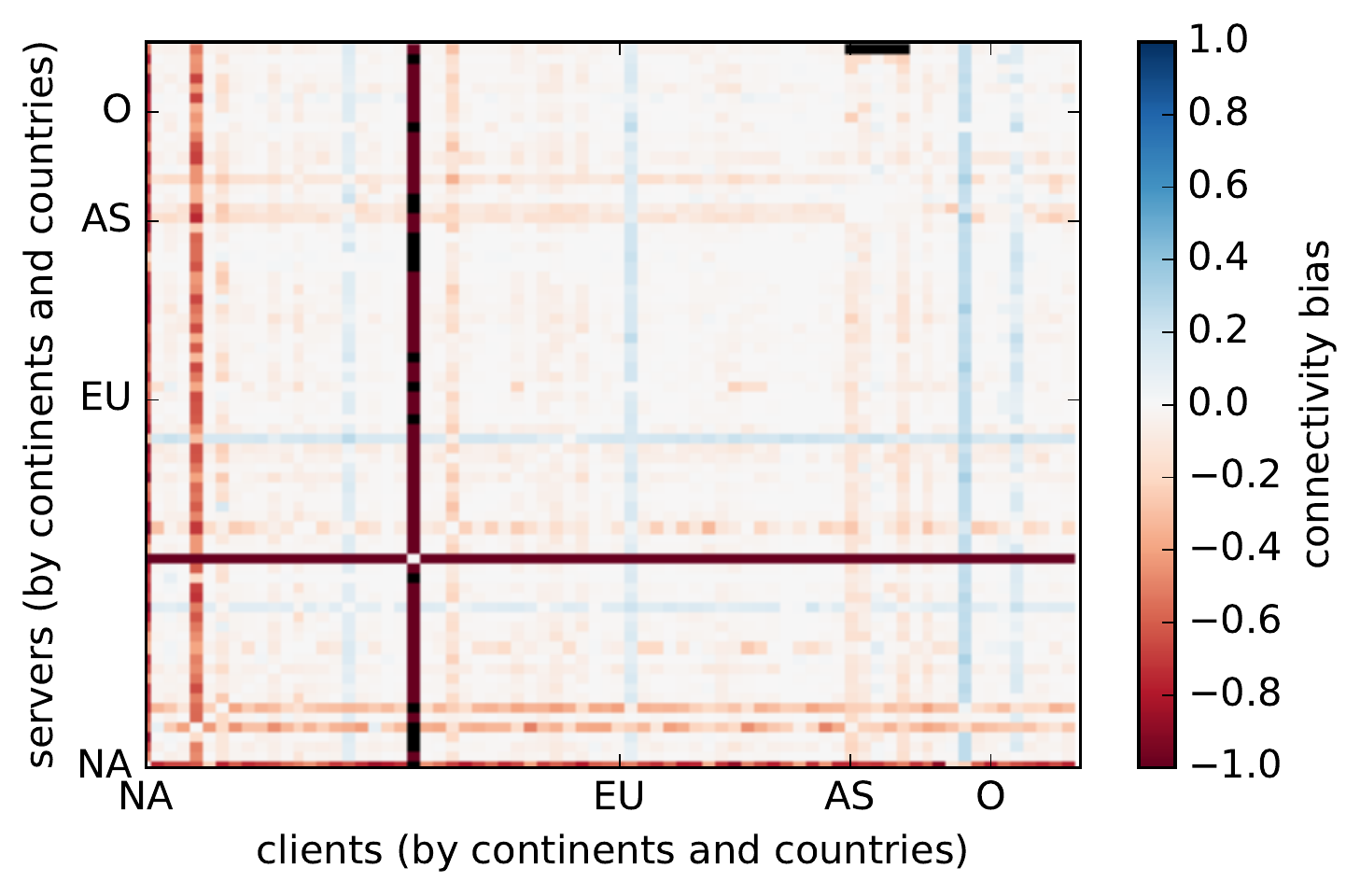}
  \end{center}  
  \caption{Connectivity bias among PlanetLab nodes, excluding ports 53 and 443.
  Positive (blue) values mean UDP is better-connected than TCP.  Black dots
  mean ``no connectivity'' (for both UDP and TCP).}
  \label{fig:conn-bias}
\end{figure}

Fig.~\ref{fig:conn-bias} shows a heatmap describing connection bias per path in
the \copycat results. A bias of +1.0 (blue) means all UDP connections between a
given receiver (X-Axis) and sender (Y-Axis) succeeded while all TCP connections
failed, and a bias of -1.0 (red) means all TCP connections succeeded while all
UDP connections failed. The axes are arranged into geographic regions: North America
(NA), Europe (EU), Asia (AS), Oceania and South America (O). The connectivity matrix
of PlanetLab nodes shown in Fig.~\ref{fig:conn-bias} confirms our findings from
Atlas, i.e., impairment is access-network linked. One node blocks all inbound
and outbound UDP traffic, and has TCP connectivity problems to some servers as well.
Otherwise, transient connectivity impairment shows a clear dependency on node,
as opposed to path.

%

\begin{table*}[!htbp]
  \begin{center}
{\small  
    \begin{tabular}{l||cc|cc||cc|cc}
    \multirow{2}{*}{\textbf{Dataset}} & \multicolumn{4}{c}{\textbf{Throughput (kB/s)}} & \multicolumn{4}{c}{\textbf{Latency (ms)}}\\
     & \multicolumn{2}{c}{\textbf{$< 200$}} & \multicolumn{2}{c}{\textbf{$> 200$}} & \multicolumn{2}{c}{\textbf{$< 50$}} & \multicolumn{2}{c}{\textbf{$> 50$}}\\
     \hline
     & \# flows & median & \# flows & median & \# flows & median & \# flows & median \\
     \hline
     PlanetLab  & 740,721 & 0.05 & 34,896 & 0.16 & 745,947 & 0.00 & 29,370 & -1.65\\
     RIPE Atlas & -     & -   & -   & -   & 2,669 & 0.00  & 48 & -4.75\\
     DO v4      & 12,563  & 0.03 & 3,637 & -0.37  & 9,381 & -0.02   & 6,819 & -0.44\\
     DO v6      & 15,459  & 0.07 & 224 & -0.16    & 15,656 & 0.00  & 27 & 3.63\\
    \end{tabular}
}    
  \end{center}
  \caption{Raw number of bias measurements (throughput and initial latency) per
  sub dataset (``DO'' stands for Digital Ocean). 
  The 50ms cut-off roughly corresponds to inter-continental versus
  intra-continental latency}
  \label{results.table.bias}
\end{table*}


\begin{table}[t]
\centering
{\small
\begin{tabular}{ r | c c }
  \hline
    \textbf{port} & \textbf{UDP blocked} & \textbf{\# probes} \\
   \hline
   53\footnote{Node pool reduced to 41 because of PlanetLab port 53 usage policies} & 0.55\% & 1,829\\   
   443\footnote{Node pool reduced to 55 because of PlanetLab port 443 usage policies} & 4.12\% & 3,034\\
   8008  & 2.60\% &  5,307\\
   12345 & 2.45\% &  5,233\\
   33435 & 2.77\% &  5,309\\
   34567 & 2.44\% &  5,115\\
   54321 & 3.07\% &  4,951\\
   \hline
\end{tabular}
}
\caption{Percentage of probes (identified as 3-tuples (IPsrc, IPdst, Portdst))
on PlanetLab, that have never seen a UDP connection but at least one TCP connection.}
\label{tab:summary_planetlab_ports}
\end{table}

\addtocounter{footnote}{-1}
\footnotetext{Node pool reduced to 41 because of PlanetLab port 53 usage policies}
\addtocounter{footnote}{1}
\footnotetext{Node pool reduced to 55 because of PlanetLab port 443 usage policies}

Table~\ref{tab:summary_planetlab_ports} shows the proportion of probes that
have never seen a UDP connection, but at least one TCP connection.  Statistics
are provided by port, as measured by \copycat on PlanetLab. As shown, UDP is
more blocked than TCP but to a small extend.  UDP is blocked in, roughly,
between 1\% and and 5\% of the probes.
We observed two China-based nodes blocking all UDP traffic from one node also
based in China. This advocates for a fall-back mechanism when running the
Internet over UDP (i.e., switching back from UDP-based encapsulation to native
TCP packets).  The absence of UDP connections for port 443
(QUIC) is mainly due to PlanetLab port binding restrictions on nodes without any
connectivity problem. Anecdotally, we found one New Zealand node blocking both
UDP and TCP traffic from all China-based nodes.

\subsection{Throughput}\label{results:throughput}

To evaluate the impact of transport-based differential treatment on throughput,
we introduce the relative $tp\_bias$ metric for each pairs of concurrent
flows.  This is computed as follows:
\begin{equation}
tp\_bias = \frac{(tp_{udp} - tp_{tcp})}{min(tp_{tcp}, tp_{udp})} \times 100 .
\label{results:throughput_bias}
\end{equation}
A positive value for $tp\_bias$ means that UDP has a higher throughput.  
A null value means that both UDP and TCP flows share the same throughput.  As no
tool able to compute throughput is available on RIPE Atlas, we only evaluate the
$tp\_bias$ on PlanetLab and Digital Ocean with \copycat.

\begin{figure}[!t]
  \begin{center}
    \includegraphics[width=8cm]{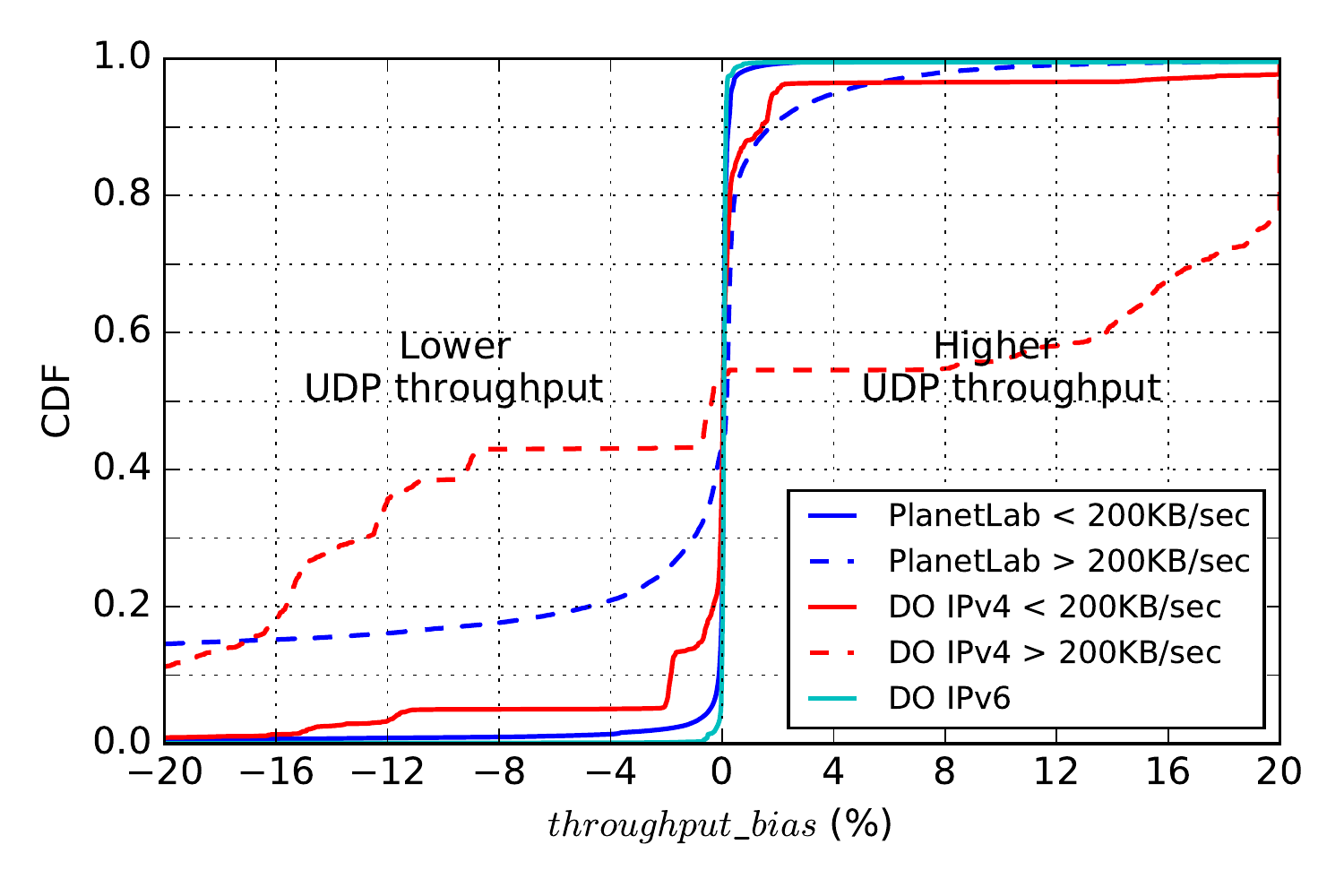}	 
  \end{center}
  \caption{Relative throughput bias, as measured by \copycat (``DO'' stands for
  Digital Ocean). Positive values mean UDP has higher throughput.  DO IPv6 has
  not been split in two due to lack of enough values (see
  Table~\ref{results.table.bias}).}
  \label{results.throughput.cdf}
\end{figure}

Fig.~\ref{results.throughput.cdf} provides a global view of the
$throughput\_bias$.  Dataset has been split between flows $<$ 200 KB/sec and
flows $>$ 200KB/sec, except for Digital Ocean IPv6, as the number of
measurements is too small to be representative. Table~\ref{results.table.bias}
gives the size of each sub dataset and the relative median bias for throughput and latency.

\begin{figure}[!t]
  \begin{center}
    \includegraphics[width=8cm]{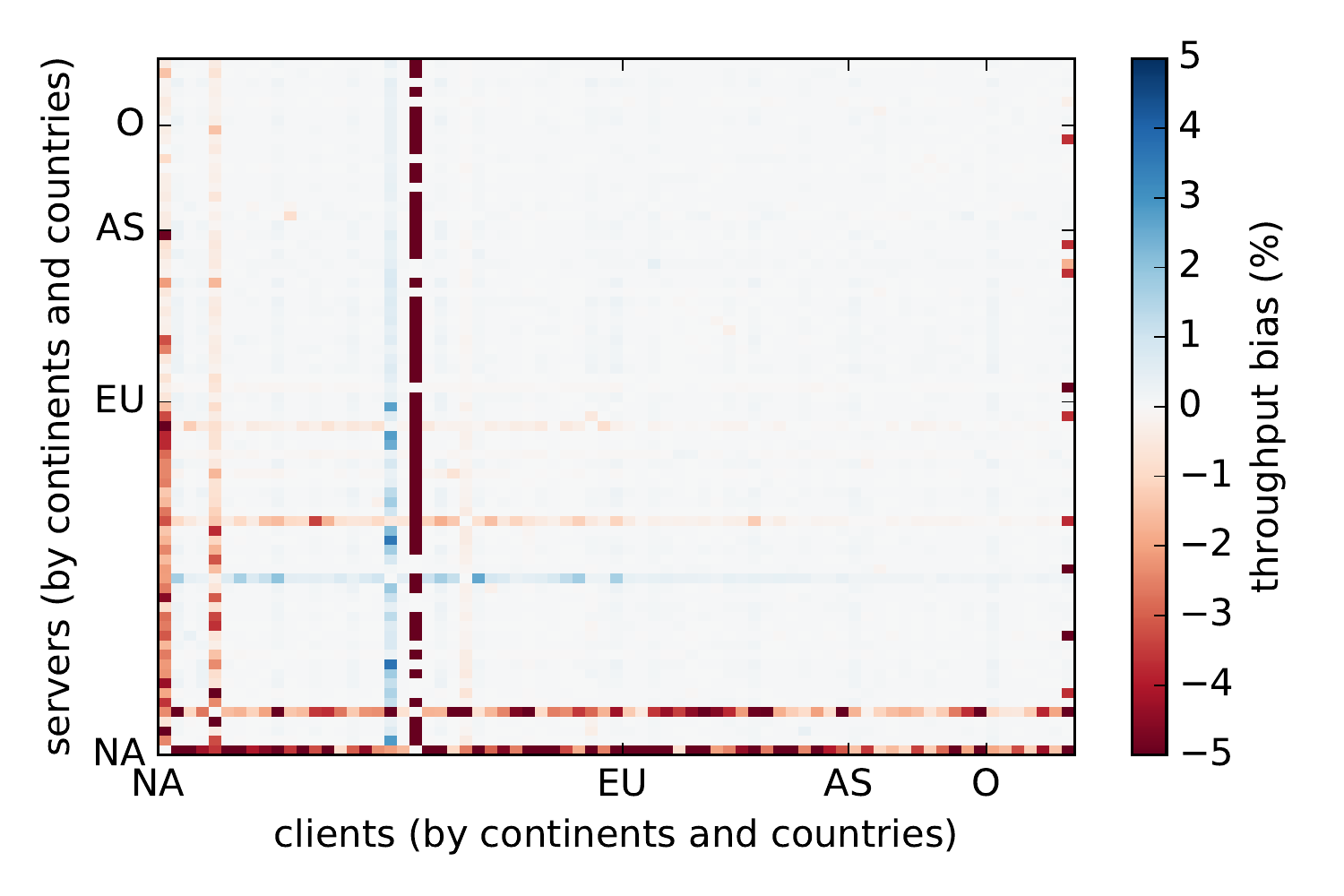}
  \end{center}
  \caption{Relative throughput bias among PlanetLab nodes as measured by
  \copycat.  Positive (blue) values mean UDP has higher throughput.}
  \label{results.throughput.grid}
\end{figure}

As observed, in general, there is no bias between UDP and TCP.  
For both Digital Ocean dataset, the non-null biases are mostly evenly
distributed in favor and disfavor of UDP. In PlanetLab, we observe an extreme
case where TCP performs better than UDP, the 4\% and 2\% highest
$throughput\_bias$ in absolute value are respectively higher than 1\% and 10\%.
As shown in Fig.~\ref{results.throughput.grid}, those extreme cases, represented
as dark red lines, are endpoint-dependent. We also notice a single  probe where
the UDP throughput is better than TCP (see Fig.~\ref{results.throughput.grid}).
Consistently with UDP connectivity bias (see Fig.~\ref{fig:conn-bias}), we do
not see evidence on path dependency for throughput.


The loss rate of congestion controlled traffic in steady state, where the link
is fully utilized, is mostly determined by the congestion control algorithm
itself. Therefore, there is a direct relation between throughput and loss.
However, as TCP congestion control reacts only once per RTT to loss as an
input signal, the actual loss rate could still be different even if similar
throughput is achieved.

Here, we understand \dfn{loss} as the percentage of flow payload lost, computed
from sequence numbers. A value, for instance, of 10\% of losses means thus that
10\% of the flow payload has been lost.

Generally speaking, the loss encountered is quite low, given that small flows
often are not large enough to fully utilize the measured bottleneck link.
As expected based on he throughput observed, we see no significant loss difference in
both PlanetLab and Digital Ocean when comparing TCP and UDP, except of 3.5\% in favor of UDP for the largest
flow size (6MB). However, this is inline with a slightly lower throughput
caused by a slightly larger initial RTT, as discussed in the next section.

\subsection{Initial Latency}\label{results:latency}


Since all \copycat traffic is congestion controlled, throughput is influenced
by the end-to-end latency. We use initial RTT measured during the TCP
handshake as a proxy for this metric.

In the fashion of $tp\_bias$ (see Eqn.~\ref{results:throughput_bias}), 
we introduce the relative $RTT\_bias$ metric for each pair of concurrent flows. 
This is computed as follows:
\begin{equation}
RTT\_bias = \frac{(RTT_{tcp} - RTT_{udp})}{min(RTT_{tcp}, RTT_{udp})} \times 100 .
\label{results:rtt_bias}
\end{equation}
A positive value for $RTT\_bias$ means that UDP has a smaller initial latency
(i.e., performs better than TCP). A null value means that both UDP and TCP flows
share the same initial latency.

\begin{figure}[!t]
  \begin{center}
    \includegraphics[width=8cm]{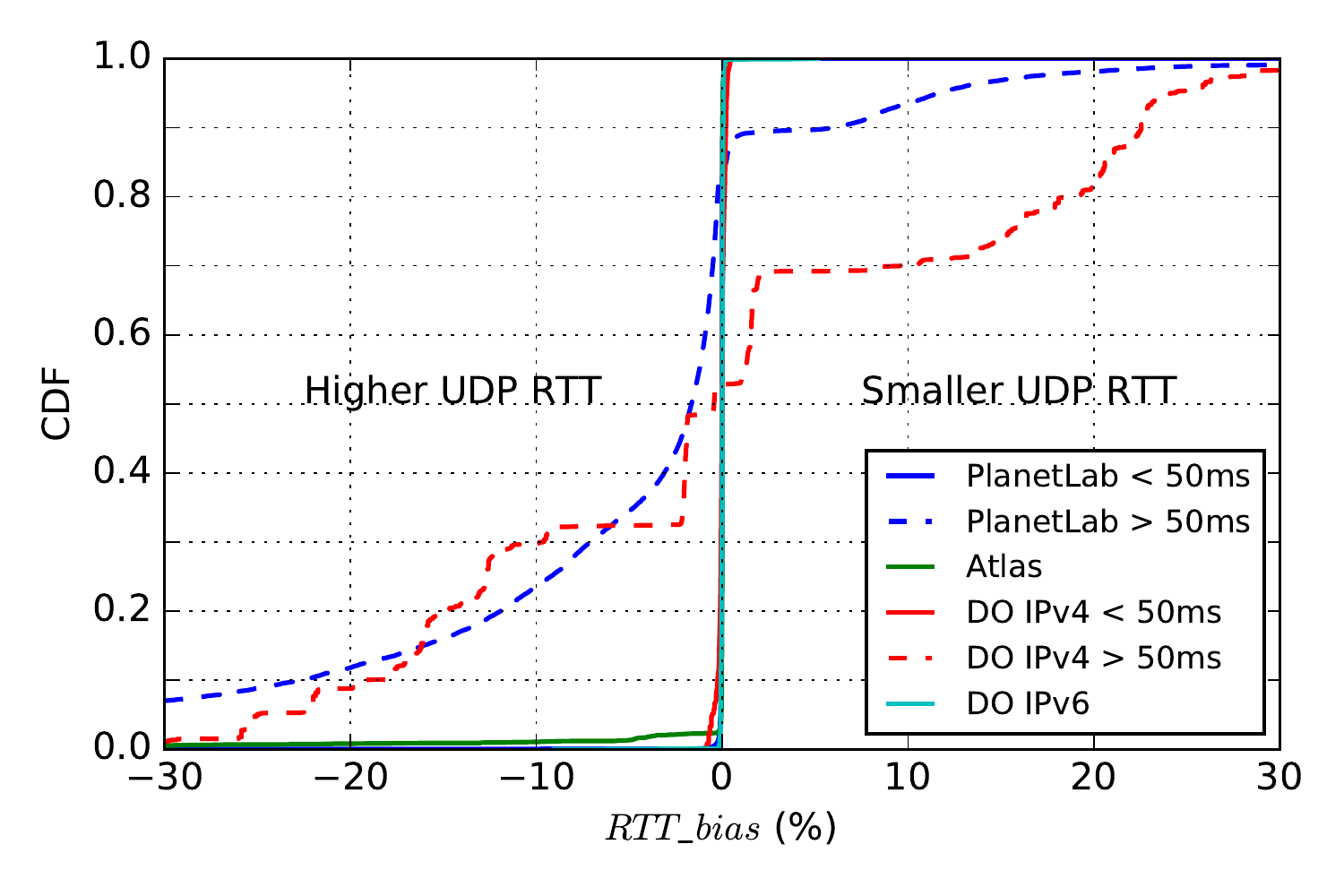}
  \end{center}
  \caption{UDP/TCP initial $RTT\_bias$ for Planetlab, RIPE Atlas (September
  2015), and Digital Ocean. Positive values mean UDP is faster.  DO IPv6 and
  Atlas have not been split in two due to lack of enough values (see
  Table~\ref{results.table.bias}).}
  \label{results:latency.irtt}
\end{figure}

The median latency bias is also listed in in Table~\ref{results.table.bias} (right part).
For PlanetLab, there is no
latency bias for flows with an initial RTT of 50ms or less, and a slight bias towards higher latency for UDP for flows with larger initial RTTs. 
For Digital Ocean we also observed a slight bias towards 
higher latency for UDP for IPv4 and no bias for IPv6 (considering 27 flows with a larger RTT than 50ms as not representative).
This is confirmed by the
CDF shown in Fig.~\ref{results:latency.irtt}.  

\begin{figure}[!t]
  \begin{center}
    \includegraphics[width=8cm]{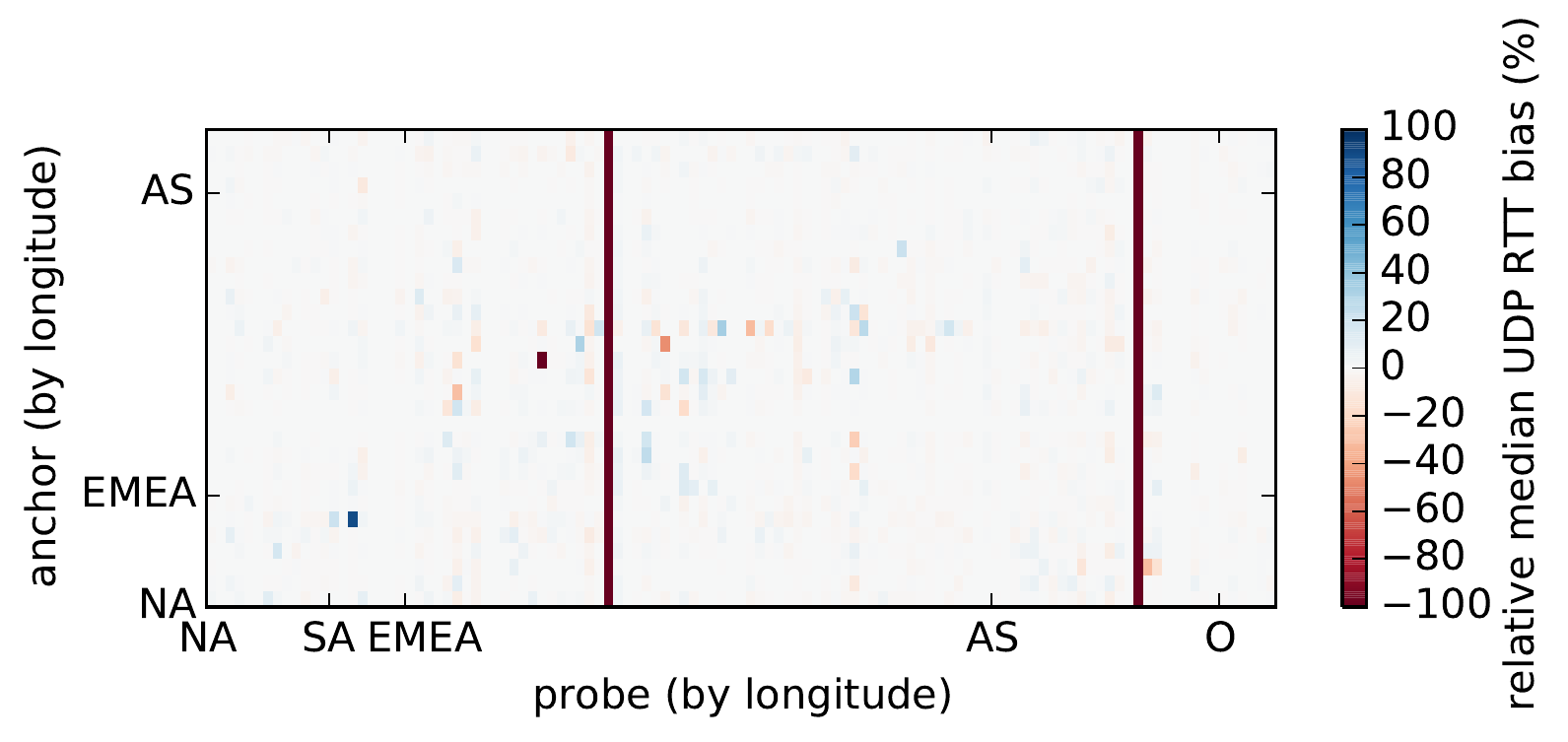}
  \end{center}
  \caption{Relative Median $RTT\_bias$, \traceroute TCP/33435 vs UDP/33435, 110 Atlas
  probes to 32 Atlas anchors, September 2015. Positive values (blue) mean UDP is
  faster.}
  \label{fig:atlas_medrtt_grid}
\end{figure}

The 2\% and 1\%  most biased flow pairs have an $RTT\_bias$ respectively lower 
than -1\% and -10\%. For the Digital Ocean IPv4 campaign, 40\% of the generated flows
have an $RTT\_bias$ between 1\% and 30\% in absolute value.
The difference between IPv4 and IPv6 on Digital Ocean appears to be due to the
presence of a middlebox interfering with all IPv4 traffic, both TCP and UDP.

This difference in latency also explains the slight throughput disadvantage as seen 
in the previous section given latency results follow nearly the same shape as 
the initial RTT (see Fig.~\ref{results:latency.irtt}).

We do not see any evidence of path dependency for initial latency in our RIPE
Atlas UDP/TCP \traceroute campaign from September 2015, either. Two probes are
on networks with a high initial latency bias toward UDP (see
Fig.~\ref{fig:atlas_medrtt_grid}), but the median over all measurements is
zero (see Fig.~\ref{results:latency.irtt}).

%
%


On PlanetLab, we see that 95\% of the flows have a latency difference of 25ms or less, and 98\% and 100\% respectively on RIPE Atlas and for IPv6 on Digital Ocean.  86\% of the flows have a difference of 10ms or less on PlanetLab, and 94\% and 87\% respectively for RIPE Atlas and for IPv6 on Digital Ocean. We consider variations of up to 10ms as usual variations in Internet measurements. Only for IPv4 we see higher variations with 87\% of flows that have a latency difference of more than 10ms  and 49\% of flows that have a latency difference of more than 25ms with -255ms for the 1\% percentile and 258ms for the 99\% percentile of the latency bias. However the median bias is still only -0.02\% which indicates that both TCP and UDP IPv4 traffic on Digital Ocean is impaired by additional in-network processing of middleboxes.

\section{Guidance and Outlook}\label{conclusion}

In this paper, we ask the question ``is UDP a viable basis and/or
encapsulation for deploying new transports in the Internet?''. We focus on two
aspects of the answer: connectivity and 
differential treatment of TCP and TCP-congestion-controlled UDP packets to see
if simply placing such traffic in UDP headers disadvantages it. Combining
these measurements with other publicly available data leads to the following
guidance for future transport protocol evolution efforts:

First, {\bf UDP provides a viable common basis} for new transport protocols,
but only {\bf in cases where alternatives exist} on access networks where UDP
connectivity is unavailable or severely compromised. QUIC provides a good
illustration here. It was developed together with SPDY, which has been defined
over TCP and TLS as HTTP/2\cite{rfc7540}, and its first target application is
HTTP/2. Since HTTP/2 has a natural fallback to TLS over TCP, this alternative
can be used on the $~1-5\%$ of networks where QUIC packets over UDP are
blocked or limited. However, this fallback approach limits QUIC's
applicability to application layer protocols that can be made run acceptably
over TCP. 

Second, our study provides evidence that the {\bf vast majority of UDP
impairments are access-network linked}, and that {\bf subtle impairment is
rare}. This means that accurate fallback decisions are easy to arrive at -- a
connection racing design similar to Happy Eyeballs~\cite{rfc6555} as used by
QUIC is sufficient -- and can often be cached based on client access network
as opposed on access-network/server pair.

However, there is still work to do. Though the study of H\"at\"onen et
al~\cite{Haetoenen2010} is six years old, the relatively mature market for
consumer-grade access points and NAT devices means that its insights are still
valid. Transports over UDP must therefore avoid NAT timeout ten to twenty
times more frequently than TCP. UDP timeout avoidance is particularly
problematic on wireless and mobile devices, since it limits the ability to
shut down the radio to reduce power consumption. Adding a generalized
mechanism for state exposure in UDP-encapsulated transport protocols is
therefore an important priority for transport evolution efforts, so that
future NAT and firewall devices can reduce the need for this additional
unproductive traffic.

We make no attempt to confirm claims of defensive rate-limiting of UDP traffic
with this work, as doing so would in essence require UDP-based denial of service
attacks on the networks under measurement. However, we note that Google reports
a reduction in the amount of UDP rate limiting they have observed since the
beginning of the QUIC experiment~\cite{quic-performance-google}. This makes
sense: rate limitation must necessarily adapt to baseline UDP traffic volumes,
and as such poses no limitation to the gradually increasing deployment of
UDP-based transport protocols. However, it also indicates the need for work on
mechanisms in these UDP protocols to augment the denial-of-service protection
afforded by rate-limiting approaches.

The authors, together with others, are working to address these issues within
the PLUS effort within the IETF~\cite{draft-trammell-plus-abstract-mech,draft-trammell-plus-statefulness},
which proposes common behavior for new UDP-encapsulated transport protocols
for managing state on devices in the network. This effort and others are
helping to overcome ossification, and will make new transport protocols
deployable within the Internet, restoring innovation in Internet services and
applications requiring transport semantics other than those provided by TCP.


\section*{Acknowledgments}
{\small
Thanks to Robert Kisteleki and Philip Homburg for assistance in applying the
RIPE Atlas network in this measurement study.

This project has received funding from the European Union's Horizon 2020
research and innovation program under grant agreement No 688421, and was
supported by the Swiss State Secretariat for Education, Research and
Innovation (SERI) under contract number 15.0268. The opinions expressed and
arguments employed reflect only the authors' views. The European Commission is
not responsible for any use that may be made of that information. Further, the
opinions expressed and arguments employed herein do not necessarily reflect
the official views of the Swiss Government.
}

\balance

\bibliographystyle{acm}
\bibliography{udp-tik}

\begin{thebibliography}{10}

\bibitem{RFC3390}
{\sc Allman, M., Floyd, s., and Partridge, C.}
\newblock Increasing {TCP}'s initial window.
\newblock {RFC} 3390, {I}nternet {E}ngineering {T}ask {F}orce, October 2002.

\bibitem{rfc7540}
{\sc Belshe, M., Peon, R., and Thomson, M.}
\newblock Hypertext transfer protocol version 2 (http/2).
\newblock {RFC} 7540, {I}nternet {E}ngineering {T}ask {F}orce, May 2015.

\bibitem{throughput_802.11}
{\sc Bruno, R., Conti, M., and Gregori, E.}
\newblock Throughput analysis and measurements in {IEEE} 802.11 {WLAN}s with
  {TCP} and {UDP} traffic flows.
\newblock {\em {IEEE} Transactions on Mobile Computing 7}, 2 (February 2008),
  171--186.

\bibitem{draft-byrne-opsec-udp-advisory}
{\sc Byrne, C., and Kleberg, J.}
\newblock Advisory guidelines for {UDP} deployment.
\newblock Internet Draft (Work in Progress) draft-byrne-opsec-udp-advisory-00,
  {I}nternet {E}ngineering {T}ask {F}orce, July 2015.

\bibitem{Carlucci2015}
{\sc Carlucci, G., De~Cicco, L., and Mascolo, S.}
\newblock Http over udp: An experimental investigation of quic.
\newblock In {\em Proceedings of the 30th Annual ACM Symposium on Applied
  Computing\/} (New York, NY, USA, 2015), SAC '15, ACM, pp.~609--614.

\bibitem{RFC6928}
{\sc Chu, J., Dukkipati, N., Cheng, Y., and Mathis, M.}
\newblock Increasing {TCP}'s initial window.
\newblock {RFC} 6928, {I}nternet {E}ngineering {T}ask {F}orce, April 2013.

\bibitem{ark}
{\sc claffy, k., Hyun, Y., Keys, K., Fomenkov, M., and Krioukov, D.}
\newblock {I}nternet mapping: from art to science.
\newblock In {\em Proc. {IEEE} Cybersecurity Applications and Technologies
  Conference for Homeland Security ({CATCH})\/} (March 2009).
\newblock see \url{http://www.caida.org/projects/ark/}.

\bibitem{digital_ocean}
{\sc {DigitalOcean}}.
\newblock Simple cloud computing, built for developers.
\newblock See \url{https://www.digitalocean.com}.

\bibitem{rfc6182}
{\sc Ford, A., Raiciu, C., Handley, M., Barre, S., and Iyengar, J.}
\newblock Architectural guidelines for multipath tcp development.
\newblock {RFC} 6182, {I}nternet {E}ngineering {T}ask {F}orce, March 2011.

\bibitem{rfc6824}
{\sc Ford, A., Raiciu, C., Handley, M., and Bonaventure, O.}
\newblock Tcp extensions for multipath operation with multiple addresse.
\newblock {RFC} 6824, {I}nternet {E}ngineering {T}ask {F}orce, Jan 2013.

\bibitem{draft-hamilton-early-deployment-quic-00}
{\sc Hamilton, R., Iyengar, J., Swett, I., and Wilk, A.}
\newblock Quic: A udp-based secure and reliable transport for http/2.
\newblock Internet-draft draft-hamilton-early-deployment-quic-00, {I}nternet
  {E}ngineering {T}ask {F}orce, July 2016.

\bibitem{Haetoenen2010}
{\sc H\"{a}t\"{o}nen, S., Nyrhinen, A., Eggert, L., Strowes, S., Sarolahti, P.,
  and Kojo, M.}
\newblock An experimental study of home gateway characteristics.
\newblock In {\em Proceedings of the 10th ACM SIGCOMM Conference on Internet
  Measurement\/} (New York, NY, USA, 2010), IMC '10, ACM, pp.~260--266.

\bibitem{natSCTP}
{\sc Hayes, D.~A., But, J., and Armitage, G.}
\newblock Issues with network address translation for {SCTP}.
\newblock {\em {ACM SIGCOMM} Computer Communication Review 39}, 1 (January
  2009), 23--33.

\bibitem{TCPExposure}
{\sc Honda, M., Nishida, Y., Raiciu, C., Greenhalgh, A., Handley, M., and
  Tokuda, H.}
\newblock Is it still possible to extend {TCP}.
\newblock In {\em Proc. {ACM} Internet Measurement Conference ({IMC})\/}
  (November 2011).

\bibitem{draft-ietf-rtcweb-data-channel}
{\sc Jesup, R., Loreto, S., and Tuexen, M.}
\newblock {WebRTC} data channels.
\newblock Internet Draft (Work in Progress) draft-ietf-rtcweb-data-channel-13,
  {I}nternet {E}ngineering {T}ask {F}orce, January 2015.

\bibitem{netalyzr}
{\sc Kreibich, C., Weaver, N., Nechaev, B., and Paxson, V.}
\newblock Netalyzr: Illuminating the edge network.
\newblock In {\em Proc. {ACM} Internet Measurement Conference ({IMC})\/}
  (November 2010).

\bibitem{draft-trammell-plus-statefulness}
{\sc Kuehlewind, M., Trammell, B., and Hildebrand, J.}
\newblock Transport-independent path layer state management.
\newblock Internet Draft (Work in Progress)
  draft-trammell-plus-statefulness-02, {I}nternet {E}ngineering {T}ask {F}orce,
  December 2016.

\bibitem{tbit}
{\sc Medina, A., Allman, M., and Floyd, S.}
\newblock Measuring interactions between transport protocols and middleboxes.
\newblock In {\em Proc. {ACM} Internet Measurement Conference ({IMC})\/}
  (November 2004).

\bibitem{handover}
{\sc Melia, T., Schmitz, R., and Bohnert, T.}
\newblock {TCP} and {UDP} performance measurements in presence of fast
  handovers in an {IP}v6-based mobility environment.
\newblock In {\em Proc. World Telecommunications Congress ({WTC})\/} (September
  2004).

\bibitem{Pahdye2001}
{\sc Pahdye, J., and Floyd, S.}
\newblock On inferring tcp behavior.
\newblock {\em SIGCOMM Comput. Commun. Rev. 31}, 4 (Aug. 2001), 287--298.

\bibitem{Paxson1997}
{\sc Paxson, V.}
\newblock End-to-end internet packet dynamics.
\newblock {\em SIGCOMM Comput. Commun. Rev. 27}, 4 (Oct. 1997), 139--152.

\bibitem{tcp_revisited}
{\sc Qian, F., Gerber, A., Mao, Z.~M., Sen, S., Spatscheck, O., and Willinger,
  W.}
\newblock {TCP} revisited: a fresh look at {TCP} in the wild.
\newblock In {\em Proc. {ACM} Internet Measurement Conference ({IMC})\/}
  (November 2009).

\bibitem{Reddy2015}
{\sc Reddy, T., Patil, P., Wing, D., and Ver~Steeg, B.}
\newblock Webrtc udp firewall traversal.
\newblock In {\em Proc. IAB Workshop on Stack Evolution in a Middlebox Internet
  (SEMI)\/} (February 2015).

\bibitem{atlasRipe}
{\sc {RIPE NCC}}.
\newblock Atlas, 2010.
\newblock See \url{https://atlas.ripe.net/}.

\bibitem{tcp_udp_queue}
{\sc Sarma, M.~P.}
\newblock Performance measurement of {TCP} and {UDP} using different queing
  algorithm in high speed local area network.
\newblock {\em International Journal of Future Computer and Communication 2}, 6
  (December 2013), 682--686.

\bibitem{rfc4960}
{\sc Stewart, R.}
\newblock Stream control transmission protocol.
\newblock {RFC} 4960, {I}nternet {E}ngineering {T}ask {F}orce, September 2007.

\bibitem{ietf96_quic}
{\sc Swett, I.}
\newblock {QUIC Deployment Experience @Google}.
\newblock https://www.ietf.org/proceedings/96/slides/slides-96-quic-3.pdf, July
  2016.

\bibitem{quic-performance-google}
{\sc Swett, I.}
\newblock {QUIC} deployment experiment @ {G}oogle, July 2016.
\newblock {IETF 96}? See
  \url{https://www.ietf.org/proceedings/96/slides/slides-96-quic-3.pdf}.

\bibitem{iperf}
{\sc Tirumala, A., Qin, F., Duagn, J., Ferguson, J., and Gibbs, K.}
\newblock {I}perf: The {TCP/UDP} bandwidth measurement tool, 2005.

\bibitem{draft-trammell-plus-abstract-mech}
{\sc Trammell, B.}
\newblock Transport-independent path layer state management.
\newblock Internet Draft (Work in Progress)
  draft-trammell-plus-abstract-mech-00, {I}nternet {E}ngineering {T}ask
  {F}orce, September 2016.

\bibitem{draft-trammell-spud-req}
{\sc Trammell, B., and Kuehlewind, M.}
\newblock Requirements for the design of a substrate protocol for user
  datagrams ({SPUD}).
\newblock Internet Draft (Work in Progress) draft-trammell-spud-req-03,
  {I}nternet {E}ngineering {T}ask {F}orce, April 2016.

\bibitem{rfc6083}
{\sc Tuexen, M., Seggelmann, R., and Rescorla, E.}
\newblock Datagram transport layer security ({DTLS}) for stream control
  transmission protocol ({SCTP}).
\newblock {RFC} 6083, {I}nternet {E}ngineering {T}ask {F}orce, January 2011.

\bibitem{rfc6555}
{\sc Wing, D., and Yourtchenko, A.}
\newblock Happy eyeballs: Success with dual-stack hosts.
\newblock {RFC} 6555, {I}nternet {E}ngineering {T}ask {F}orce, April 2012.

\bibitem{Xu2014}
{\sc Xu, Y., Wang, Z., Leong, W.~K., and Leong, B.}
\newblock An end-to-end measurement study of modern cellular data networks.
\newblock In {\em Proceedings of the 15th International Conference on Passive
  and Active Measurement - Volume 8362\/} (New York, NY, USA, 2014), PAM 2014,
  Springer-Verlag New York, Inc., pp.~34--45.

\end{thebibliography}

\section*{Appendix: Repeatability}
The measurements and results discussed in this paper are intended to be
repeatable by others in the Internet measurement and transport protocol
research community. This appendix discusses how.

\subsection*{Copycat}
\copycat, as presented in Section \ref{methodology.copycat} is freely available at 
\url{https://github.com/mami-project/udptun}.  The git repository contains a
\texttt{README.md} file that fully explains how to compile, deploy, and use
\copycat.

\subsection*{Data and Analysis}
The dataset collected and analyzed (PlanetLab, RIPE Atlas, and Digital Ocean) in
this paper is freely available at \url{https://github.com/mami-project/udpdiff}.

Data analysis was performed using Python and Pandas; Jupyter notebooks for
performing analysis done in this paper are available along with the datasets
at \url{https://github.com/mami-project/udpdiff}.

\end{document}